\documentclass{ws-mpla}
\usepackage[super]{cite}
\usepackage{graphicx}

\def\kb{k_{B}}
\def\F{\mathcal{F}}
\def\rmd{\mathrm{d}}
\def\rme{\mathrm{e}}
\def\tm{\mathrm{TM}}
\def\te{\mathrm{TE}}

\begin{document}

\markboth{C.~C.~Korikov \& V.~M.~Mostepanenko}
{Nernst Heat Theorem for the~Casimir-Polder Interaction}

\catchline{}{}{}{}{}

\title{\uppercase{Nernst heat theorem for the~Casimir-Polder interaction between a magnetizable atom and ferromagnetic dielectric plate}}

\author{\uppercase{C.~C.~Korikov$^1$ {\lowercase{ and}} V.~M.~Mostepanenko$^{1,2,3}$}}

\address{$^1$Institute of Physics, Nanotechnology and Telecommunications,
Peter the Great\\ Saint Petersburg Polytechnic University,
Saint Petersburg, 195251, Russia\\
$^2$Central Astronomical Observatory at Pulkovo of the Russian Academy of Sciences,\\
Saint Petersburg, 196140, Russia\\
$^3$Kazan Federal University, Kazan, 420008, Russia\\
constantine.korikov@gmail.com,  vmostepa@gmail.com}

\maketitle

 \pub{Received 12 July 2019}{Revised 8 August 2019}

\begin{abstract}
We find the low-temperature behavior of the Casimir-Polder free energy for a polarizable and magnetizable atom interacting with a~plate made of
ferromagnetic dielectric material. It is shown that the corresponding
Casimir-Polder entropy goes to zero with vanishing temperature,
i.e., the Nernst heat theorem is satisfied, if the dc conductivity of the plate
material is disregarded in calculations. If the dc conductivity is taken into account, the Nernst theorem is violated.
These results are discussed in light of recent experiments.

\end{abstract}

\ccode{PACS Nos.: 12.20.Ds, 34.35.+a, 65.40.gd}

\section{Introduction}	
The Nernst heat theorem has long been used as a test for different
theoretical approaches to the Casimir force. Thus, in Ref.~\refcite{ref1}
the behavior of the Casimir free energy and entropy at low temperature was found in the configuration of two parallel ideal metal plates.
It was proven that at zero temperature the Casimir entropy is equal to zero in accordance with the Nernst theorem. At a later time, it was shown that for two
metallic plates with perfect crystal lattice the Casimir entropy
satisfies the Nernst theorem if the low-frequency dielectric response of metal is described
by the lossless plasma model and violates this theorem if the
lossy Drude model is used~\cite{ref2,ref3,ref4,ref5}. The same is true
for a metallic sphere and a plate~\cite{ref6} and for two plates made of
ferromagnetic metal~\cite{ref7}. For an atom interacting with metallic plate the Nernst theorem
is followed regardless of the used model of dielectric response of metal
if this atom does not possess magnetic properties. For a~magnetizable atom, however,
the Casimir-Polder entropy satisfies and violates the Nernst theorem when
the plasma and Drude models are used, respectively, like for two metallic
plates~\cite{ref8}.

For two dielectric plates described with omitted conductivity at a constant current (dc conductivity), the Casimir entropy satisfies the Nernst theorem and violates
it otherwise~\cite{ref9,ref10,ref11}. The same holds for two plates made of ferromagnetic dielectric~\cite{ref12}
and for the Casimir-Polder entropy of a polarizable atom interacting with nonmagnetic dielectric plate~\cite{ref13}.
Thus, for a~dielectric plate, thermodynamic properties
depend on the used model of dielectric response even if an atom is not magnetizable.

In this paper, we find an analytic expression for the Casimir-Polder
free energy at low temperature in more general
case of polarizable and magnetizable atom interacting
with a plate made of ferromagnetic dielectric.
We show that the respective Casimir-Polder
entropy satisfies the Nernst theorem if the dc conductivity of a plate is omitted in calculations and violates
it if the dc conductivity is included.

\section{The Casimir-Polder Free Energy at Low Temperature}
We start with the Lifshitz formula for the free energy of Casimir-Polder interaction between
both polarizable and magnetizable atom and a ferromagnetic dielectric plate
expressed in terms of dimensionless variables~\cite{ref14}
\begin{equation}\label{eq1}
\F(a, T) = \frac{\kb T}{8 a^3} \sum_{l=0}^{\infty}\phantom{}^{'} \Phi(\zeta_l).
\end{equation}
Here, $\kb$ is the Boltzman constant, $T$ is temperature of the plate
and the environment, $a$ is a distance between the ground state atom and the
plate. The prime on the summation sign divides the term $l=0$ by 2 and the function $\Phi$ is defined~as
\begin{equation}\label{eq2}\Phi(\zeta_l) = \Phi^{(\alpha)}(\zeta_l) + \Phi^{(\beta)}(\zeta_l),
\end{equation}
where the dimensionless Matsubara frequencies $\zeta_l$ are expressed
via the dimensional ones as $\zeta_l=\xi_l/\omega_{c}$, the characteristic
frequency is $\omega_c=c/(2a)$ and
\begin{equation}
\begin{split}
    \Phi^{(\alpha)}(\zeta_l) &= -\alpha_l \int_{\zeta_l}^{\infty} \rmd y \;\rme^{-y} \left[2y^2 r_{\tm,l}-\zeta_l^2(r_{\tm,l}+r_{\te,l})\right],\\
    \Phi^{(\beta)}(\zeta_l) &= -\beta_l \int_{\zeta_l}^{\infty} \rmd y \;\rme^{-y} \left[2y^2 r_{\te,l}-\zeta_l^2(r_{\tm,l}+r_{\te,l})\right].
\end{split}
\label{eq3}
\end{equation}
In these equations,  $\alpha_l=\alpha(i\omega_{c}\zeta_l)$ and $\beta_l=\beta(i\omega_{c}\zeta_l)$ are the atomic
electric and magnetic polarizabilities. The reflection coefficients for the transverse magnetic (TM) and transverse electric (TE)
polarizations of the electromagnetic field are defined~as
\begin{equation}
\begin{split}
    r_{\tm,l} = r_{\tm}(i\zeta_l,y) &= \frac{\epsilon_l y - \sqrt{y^2+\zeta_l^2\gamma_l}}{\epsilon_l y + \sqrt{y^2+\zeta_l^2\gamma_l}},\\
    r_{\te,l} = r_{\te}(i\zeta_l,y) &= \frac{\mu_l y - \sqrt{y^2+\zeta_l^2\gamma_l}}{\mu_l y + \sqrt{y^2+\zeta_l^2\gamma_l}},
\end{split}
\label{eq4}
\end{equation}
where $\gamma_l=\epsilon_l\mu_l-1$, $\epsilon_l=\epsilon(i\omega_c \zeta_l)$ and $\mu_l=\mu(i\omega_c \zeta_l)$ are the dielectric
permittivity and magnetic permeability of ferromagnetic dielectric plate at the ima\-gi\-nary Matsubara frequencies. Note that ferromagnetic
dielectrics possess physical properties characteristic for
dielectrics (i.e., low conductivity permitting the use of finite values of $\epsilon_0$) but behave like ferromagnets in
an external magnetic field~\cite{ref15,ref16}.

By means of the Abel-Plana formula~\cite{ref14}, the temperature-dependent contribution to the free energy~\eqref{eq1} can be
written in the form
\begin{equation}\label{eq5}
    \Delta\F(a,T) = \frac{i\kb T}{8 a^3} \int_{0}^{\infty} \rmd t \; \frac{\Phi(i \tau t)-\Phi(-i\tau t)}{\rme^{2\pi t}-1},
\end{equation}
where in the above formulas $\zeta_l=4\pi a \kb T l/(\hbar c) \equiv \tau l$ should be replaced with $\tau t$.

Now we disregard the dc conductivity of the plate material and consider the frequency-independent permittivity and
permeability $\epsilon_l=\epsilon_0$, $\mu_l=\mu_0$. The latter assumption increases by one the power of the primary contribution to
the low-temperature asymptotic expansion of the Casimir-Polder free energy~\cite{ref12,ref17}.
We also put $\alpha_l=\alpha_0$ and $\beta_l=\beta_0$ which does not influence the primary contribution of this expansion~\cite{ref13}.
By putting $\tau t = x$ and introducing the new integration variable $z=y/x$ in Eq.~\eqref{eq3},
we can rewrite Eq.~\eqref{eq2}~as
\begin{equation}\label{eq6}
    \Phi(x)=-\alpha_0 I(x,\epsilon_0)-\beta_0 I(x,\mu_0),
\end{equation}
where
\begin{equation}\label{eq7}
    I(x,\eta)=x^3 \int_1^\infty \rmd z \; \rme^{-x z} \left[2 z^2 r(\eta, z)-r(\epsilon_0,z)-r(\mu_0,z)\right]
\end{equation}
and
\begin{equation}\label{eq8}
    r(\eta,z)=\frac{\eta z-\sqrt{z^2+\gamma_0}}{\eta z + \sqrt{z^2+\gamma_0}}.
\end{equation}
In order to make integrals convergent after putting $x=0$ in the power of an exponent,
we subtract two necessary terms from $r(\eta,z)$ and by one from $r(\epsilon_0,z)$ and $r(\mu_0,z)$.
To preserve an equality, the same terms are added. As a result, Eq.~\eqref{eq7} takes the form
\begin{equation}\label{eq9}
    I(x,\eta) = x^3 \left[I_1(x,\eta)+I_2(x)+I_3(x,\eta)\right],
\end{equation}
where
\begin{equation}
\begin{split}
    I_1(x,\eta) &= 2 \int_1^\infty \rmd z  \;\rme^{-x z}\left\{z^2\left[r(\eta,z)-\frac{\eta-1}{\eta+1}\right]+\frac{\eta \gamma_0}{(\eta+1)^2}\right\},\\
    I_2(x) &=\int_1^\infty \rmd z \; \rme^{-x z}  \left[-r(\epsilon_0,z)+\frac{\epsilon_0-1}{\epsilon_0+1}-r(\mu_0,z)+\frac{\mu_0-1}{\mu_0+1}\right],\\
    I_3(x,\eta) &= \int_1^\infty \rmd z \;\rme^{-x z} \left[2 z^2 \frac{\eta-1}{\eta+1}-A(\eta)\right]
\end{split}
\label{eq10}
\end{equation}
and the following notation is introduced:
\begin{equation}\label{eq11}
    A(\eta) = \frac{2 \eta \gamma_0}{(\eta+1)^2} + \frac{\epsilon_0-1}{\epsilon_0+1} + \frac{\mu_0-1}{\mu_0+1}.
\end{equation}
Calculating $I_3$ precisely, for the first terms of expansion in powers of $x$ one obtains
\begin{equation}\label{eq12}
    x^3 I_3(x,\eta) = 4 \frac{\eta-1}{\eta+1}-A(\eta)x^2+\frac{1}{3}\left[3A(\eta)-2\frac{\eta-1}{\eta+1}\right]x^3.
\end{equation}
Calculating $I_3$ precisely, for the first terms of expansion in powers of $x$ one obtains
\begin{equation}\label{eq12}
    x^3 I_3(x,\eta) = 4 \frac{\eta-1}{\eta+1}-A(\eta)x^2+\frac{1}{3}\left[3A(\eta)-2\frac{\eta-1}{\eta+1}\right]x^3.
\end{equation}
The first expansion terms of $I_1$ and $I_2$ are of the order $x^{0}$ and are obtained from Eq.~\eqref{eq10} by putting $x=0$ in the
powers of exponents using the following integrals:
\begin{eqnarray}
&&
    \int \rmd z \; \frac{\eta z - \sqrt{z^2+\gamma_0}}{\eta z + \sqrt{z^2+\gamma_0}} =
    \frac{1}{\eta^2-1} \left[(\eta^2+1)z-2\eta\sqrt{z^2+\gamma_0}\right.
\nonumber \\
&&\left.-\frac{2\eta^2\sqrt{\gamma_0}}{\sqrt{\eta^2-1}}\ln{\frac{\sqrt{\gamma_0}+
z\sqrt{\eta^2-1}}{\eta\sqrt{\gamma_0}+\sqrt{z^2+\gamma_0}\sqrt{\eta^2-1}}}\right] + C,
\label{eq13}\\
    &&
\int \rmd z \; z^2\frac{\eta z - \sqrt{z^2+\gamma_0}}{\eta z + \sqrt{z^2+\gamma_0}} =
    \frac{2}{3(\eta^2-1)^2}
\left\{\vphantom{\frac{3\eta^2\gamma_0^{3/2}}{\sqrt{\eta^2-1}}}
3 \gamma_0 \eta^2 z + \frac{1}{2} \left(\eta^4 -1 \right) z^3
-\eta \sqrt{z^2 + \gamma_0}\right.
\nonumber\\
    &&\times \left[(\eta^2-1)z^2+\gamma_0\left(\eta^2+2\right)\right]
\left.-\frac{3\eta^2\gamma_0^{3/2}}{\sqrt{\eta^2-1}}\ln{\frac{\sqrt{\gamma_0}+z\sqrt{\eta^2-1}}{\eta\sqrt{\gamma_0}+\sqrt{z^2+\gamma_0}\sqrt{\eta^2-1}}}\right\}+C.
\nonumber
\end{eqnarray}
As a result, for $\Phi(x)$ defined in Eq.~\eqref{eq6} we find
\begin{equation}
\begin{split}
    \Phi(x) = -\alpha_0& \left[4\frac{\epsilon_0-1}{\epsilon_0+1}-A(\epsilon_0)x^2+B_\alpha(\epsilon_0,\mu_0)x^3\right]\\
              -\beta_0&\left[4\frac{\mu_0-1}{\mu_0+1}-A(\mu_0)x^2+B_\beta(\epsilon_0,\mu_0)x^3\right],
\end{split}
\label{eq14}
\end{equation}
where
\begin{eqnarray}
&&
    B_\alpha(\epsilon_0,\mu_0) = \frac{\mu_0^2+1}{\mu_0^2-1} +
    \frac{\epsilon_0^4-12\epsilon_0^2\gamma_0-1}{3(\epsilon_0^2-1)^2}
+\frac{2\epsilon_0^2\sqrt{\gamma_0}(2\epsilon_0\mu_0-1-\epsilon_0^2)}{(\epsilon_0^2-1)^{5/2}}
\nonumber\\
    &&
\times
\ln{\frac{\sqrt{\gamma_0}+\sqrt{\epsilon_0^2-1}}{\epsilon_0\sqrt{\gamma_0}+
\sqrt{\epsilon_0\mu_0}\sqrt{\epsilon_0^2-1}}}
-2\sqrt{\epsilon_0\mu_0}
    \left[
        \frac{\epsilon_0}{3\left(\epsilon_0^2-1\right)}+
        \frac{\mu_0}{\mu_0^2-1}-
        \frac{2}{3}\frac{\epsilon_0(\epsilon_0^2+2)\gamma_0}{(\epsilon^2-1)^2}
    \right]
\nonumber\\
&&
    -\frac{2\mu_0\sqrt{\gamma_0}}{(\mu_0^2-1)^{3/2}}
\ln{\frac{\sqrt{\gamma_0}+\sqrt{\mu_0^2-1}}{\mu_0\sqrt{\gamma_0}+
\sqrt{\epsilon_0\mu_0}\sqrt{\mu_0^2-1}}}
\label{eq15}
\end{eqnarray}
and $B_\beta(\epsilon_0,\mu_0)$ is obtained from $B_\alpha(\epsilon_0,\mu_0)$ by interchanging $\epsilon_0$ and $\mu_0$.

Substituting~\eqref{eq14} in~\eqref{eq5} and integrating, one arrives at
\begin{equation}\label{eq16}
    \Delta \F(a,T) = - \frac{\pi^3 (\kb T)^4}{15 (\hbar c)^3} \left[\alpha_0 B_\alpha(\epsilon_0,\mu_0)+\beta_0B_\beta(\epsilon_0,\mu_0)\right].
\end{equation}
It is seen that~\eqref{eq16} does not depend on the atom-plate separation.
Note that in the limiting case $\beta_0=0, \mu\to1$ Eq.~\eqref{eq15} reduces to
\begin{eqnarray}
&&    B_{\alpha}(\epsilon_0,1)=\frac{\epsilon_0-1}{\epsilon_0+1}\frac{7\epsilon_0+1}{3(\epsilon_0+1)}
 -\frac{2\epsilon_0^2}{(\epsilon_0+1)^{5/2}}
    \ln{\frac{1+\sqrt{\epsilon_0+1}}{\sqrt{\epsilon_0}(\sqrt{\epsilon_0}+\sqrt{\epsilon_0+1})}}
\nonumber\\
    &&+\frac{(\sqrt{\epsilon_0}-1)\left[(3\epsilon_0^2+1)(2\sqrt{\epsilon_0}+1)+
    2\epsilon_0(\sqrt{\epsilon_0}-1)\right]}{3(\sqrt{\epsilon_0}+1)(\epsilon_0+1)^2}.
\label{eq17}
\end{eqnarray}
This coincides with Eq.~(7) in Ref.~\refcite{ref13} except that the factor $2\epsilon_0$ in front
of the last term in square brackets was indicated in Ref.~\refcite{ref13} in error as
$3\epsilon_0$.

\section{The Nernst Heat Theorem and the dc Conductivity}
From~Eq.~\eqref{eq16} one can calculate the Casimir-Polder entropy at low temperature
\begin{equation}\label{eq18}
    S(T)=-\frac{\partial \Delta \F(T)}{\partial T} = \frac{4\pi^3}{15(\hbar c)^3} \kb (\kb T)^3 \left[\alpha_0 B_\alpha (\epsilon_0,\mu_0)+\beta_0 B_\beta(\epsilon_0,\mu_0)\right].
\end{equation}
Thus, $S(0)=0$, i.e., for an ideal ferromagnetic dielectric plate with $\epsilon_0 < \infty$ and a magnetizable atom
the Nernst heat theorem is satisfied. Therefore the Lifshitz theory in this configuration  is thermodynamically consistent,
provided that the conductivity of ferromagnetic dielectric material at a constant current is disregarded.

Taking into consideration that conductivity of dielectrics, while very small, is really existing effect,
one may ask what is its impact on the Casimir-Polder entropy. With account of conductivity
the permittivity of plate material $\epsilon_l$ is replaced by~\cite{ref14}
\begin{equation}\label{eq19}
    \widetilde{\epsilon}_l = \epsilon_l + \frac{4\pi \widetilde{\sigma}_0(T)}{\zeta_l},
\end{equation}
where the dimensionless conductivity is connected with dimensional by $\tilde{\sigma}_0=\sigma_0/\omega_c$.

It was shown~\cite{ref9} that the transition from $\epsilon_l$ to $\widetilde{\epsilon}_l$ leads to only exponentially
small in $T$
additions to all terms of the Lifshitz formula with $l \ge 1$. As to the TM reflection coefficient at $l=0$, it
 takes the values
\begin{equation}\label{eq20}
    r_{\tm,0} = \frac{\epsilon_0-1}{\epsilon_0+1}, \;\; \widetilde{r}_{\tm,0}=1
\end{equation}
when the permittivities $\epsilon_l$ and $\widetilde{\epsilon}_l$ are substituted in
Eq.~\eqref{eq4}, respectively. In so doing, the reflection coefficient $r_{\te,0}$ remains unchanged. Because of this, up to
the terms decaying to zero exponentially fast with vanishing temperature, one obtains
\begin{equation}\label{eq21}
    \widetilde{\F}(a,T) = \F(a,T) - \frac{\kb T}{4 a^3} \left(1-\frac{\epsilon_0-1}{\epsilon_0+1}\right)\alpha_0,
\end{equation}
where $\F$ is found with the dc conductivity
disregarded. From this equation we find
\begin{equation}\label{eq22}
    \widetilde{S}(a,0) = \frac{\kb \alpha_0}{2 a^3}\frac{1}{\epsilon_0+1}>0.
\end{equation}
Thus, the Casimir-Polder entropy at zero temperature is equal to the positive quantity depending on the parameters
of a system (static atomic polarizability, permittivity and atom-plate separation), i.e., the Nernst heat theorem is violated.

\section{Conclusions and Discussion}
We have found the asymptotic behavior at low $T$
of the Casimir-Polder free energy and entropy for a polarizable and magnetizable atom interacting with a ferromagnetic
dielectric plate. It is shown that the Nernst theory is satisfied if the dc
conductivity of plate material is disregarded and is
violated otherwise.

It is of prime importance that the experimental data for both Casimir
and Casimir-Polder forces are in agreement with thermodynamically consistent theoretical approach disregarding dc
conductivity of dielectric materials (see review in Ref.~\refcite{ref14} and more recent results~\cite{ref18,ref19}). In this respect,
it should be mentioned that for graphene systems, where the dielectric response is found
basing on the first principles of quantum electrodynamics,
the Casimir and Casimir-Polder entropy satisfies the Nernst theorem~\cite{ref20,ref21}.
To conclude, thermodynamic problems, arising in application of the Lifshitz theory to real materials,
invite further investigation.

\section*{Acknowledgments}

V.~M.~Mostepanenko was partially funded by the Russian Foundation for Basic Research,
Grant No. 19-02-00453~A. His work was also partially supported by the Russian Government Program of
Competitive Growth of Kazan Federal University.

\end{document}